\documentclass[eqsecnum,aps,epsfig]{revtex4}
\usepackage{amsmath,amssymb,amsthm,bm}
\usepackage{psfrag}
\usepackage[dvips]{graphicx}

\begin{document}
\title{Confinement in G(2) Gauge Theories Using Thick Center Vortex Model and domain
structures}
\author{S.~Deldar$^1$ and H.~Lookzadeh$^2$ and S.M.~Hosseini~Nejad$^3$}
\affiliation{
Department of Physics, University of Tehran, P.O. Box 14395/547, Tehran 1439955961,
Iran \\
$^1$sdeldar@ut.ac.ir\\
 $^2$h.lookzadeh@ut.ac.ir\\
 $^3$smhosseini211@gmail.com
 }
\date{\today}

\begin{abstract}

The thick center vortex model with the idea of using domain structures is used to
calculate the potentials between two G($2$) heavy sources in the fundamental, the
adjoint and the $27$ dimensional representations. The potentials are screened at
large distances. This behavior is expected from the thick center vortex model since
G($2$) has only a trivial center element which makes no contribution to the average
Wilson loop at the large distance regime. A linear potential is obtained at
intermediate distances for all representations. This behavior can be explained by
the thickness of the vortices (domains) and by defining a flux for the trivial
center element of G($2$). The role of the SU($3$) subgroup of G($2$) in the linear
regime is also discussed. The string tensions are in rough agreement with the
Casimir operators of the corresponding representations.
\end{abstract}

\maketitle

\section{INTRODUCTION}

The last candidate for the fundamental theory of hadronic force is the quark model.
The interaction between quarks is described by the exchange of the non-Abelian gauge
fields called gluons.
Quarks were introduced by Gellman and Zweig for the first time in 1964.
Besides the successfulness of the quark model, no free quark has been observed yet.
However, the particles in the Lagrangian of any theory must exist in the physical
spectrum.
Now that the quarks appear in the QCD Lagrangian, the main question is as follows: Where are
the quarks? Are they permanently confined?
In fact, these are not the quarks which are not found in the nature but it is the
color which is confined and quarks are confined because of their colors. In other
words, particles with color degrees of freedom are confined. Therefore, all free
particles are colorless, and colored particles cannot be found in the particle
spectrum. Also gluons, particles in the adjoint representation of the gauge group of
QCD have colors, and are absent in the particle spectrum, as well.

\begin{figure}[b]
\begin{center}
\resizebox{0.87\textwidth}{!}{
\includegraphics{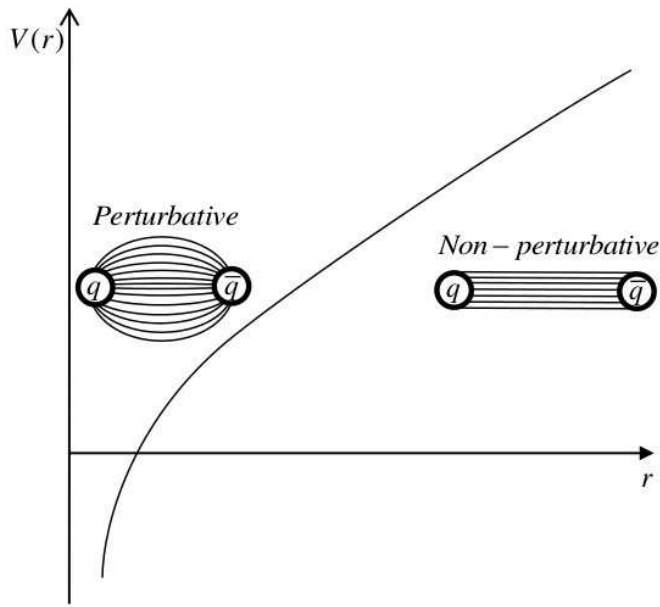}}
\caption{\label{pot}
Potential between static quarks. At small distances the coupling constant is small
enough to use perturbative methods. A Coulombic potential is obtained in this
regime. At large distances the coupling constant is large, and nonperturbative
effects lead to a linear potential.}
\end{center}
\end{figure}

Let us define confinement from the point of view of the potential in a
quark-antiquark pair. At short distances the potential in a quark-antiquark
pair is Coulombic and can be explained by the gluon exchange between quarks. Since the
coupling constant is small, perturbative methods work very well in this region. As
the distances between quarks increase, the coupling constant increases as well and
perturbative techniques do not work anymore. This happens at an intermediate regime
where the potential between quarks rises linearly until the energy reaches a point
where it is large enough to create a quark-antiquark pair in the vacuum. Then, these
secondary pairs couple to the initial ones and screen them. However, by removing the
dynamical quarks from the theory, the creation of the secondary quarks does not
happen and the potential between the initial quarks rises linearly with distance.
Figure (\ref{pot}) shows a typical behavior of the potential between static quark
 s, a Coulombic potential at short distances, and a linear potential for intermediate
and large distances. For this figure, dynamical quarks are removed from the theory.
The linear behavior is confirmed by Regge trajectories in the particle spectrum
\cite{Collins1968}. The flux tube created at this linear regime has nonlinear
nature, and it is not possible to use ordinary perturbation methods to study its
characteristics.
Therefore, to investigate the quark confinement phenomena, one should use
nonperturbative methods. Lattice gauge theory has been very successful to reproduce
the potential between pairs of colored particles like quarks and gluons. Other
approaches used to study confinement include phenomenological models. In these models, the
QCD vacuum is filled with some topological configurations confining colored objects.
The most popular candidates among these topological fields are monopoles and
vortices. Other candidates include merons, calorons,$\it{etc}$. There are very
strong correlations between these various objects, though. Therefore, the mechanism
of confinement has not been understood by a unique procedure yet, and it is one of
the unsolved mysteries of quantum chromodynamics.

In this paper, we study the center vortex model within the G(2) gauge group. The
center vortex model was initially introduced by 't Hooft \cite{Hooft1979}. It was
able to explain the confinement of quark pairs, but it was not able to explain
the intermediate linear potential,especially for higher representations.The model
has been improved to the thick center vortex model by M. Faber $\it{et~ al.}$
\cite{Fabe1998}. Even though it is a simple model, the thick center vortex model
reproduces the intermediate linear potential qualitatively, in agreement with Casimir
scaling and the asymptotic large distance potential which is proportional to the $N$-ality of
the gauge group. For intermediate distances, one expects to see a linear potential
proportional to the distance between quarks, in agreement with lattice calculations.
Based on lattice results \cite{Bali2000,Deld1999}, the coefficient of the linear
part which is called the string tension, is proportional to the eigenvalue of t
 he quadratic Casimir operator of the
corresponding representation. This is called Casimir scaling. The Casimir scaling
regime is expected to extend roughly from the onset of confinement to the onset of
screening. The $N$-ality of each representation is given by mod($n-m$), where $n$ is
the number of quarks and $m$ is the number of antiquarks constructing the
representation. At large distances, where the distance between the quark and antiquark
increases, the energy stored in the string increases and gluon pairs are
created from the vacuum. These gluons screen the initial sources to the lowest
dimension with the same $N$-ality. In this regime, sources with the same $N$-ality
obtain the same string tensions. Those with zero $N$-ality are completely screened.

In the vortex model the QCD vacuum is filled with linelike (in three dimensions) or
surfacelike (in four dimensions) objects, which carry magnetic flux quantized in
terms of the center elements of the gauge group. At large distances between color
sources, the interactions between Wilson loops and center vortices lead to a linear
potential proportional to the $N$-ality of the given representation.

Besides the successes of the center vortex model, the relation between confinement
and center symmetry has not been understood very well,yet. To study the role of
center elements, one of the attractive methods is studying confinement in gauge
groups without nontrivial center elements, like G($2$). Lattice calculations in
G($2$) \cite{Olejnik2008} show color screening for large distances and a linear
potential for intermediate distances. Since the center is trivial, the center vortex
model predicts color screening. It is not so clear why a linear potential can be
expected at intermediate distances. Because of these two features confinement at
intermediate distances and screening at large distances G(2) gauge theory resembles
SU($N$) gauge theories with dynamical quarks. These screen static quarks at large
distances. Therefore, G($2$) is considered as a mathematical laboratory to examine
the properties of SU($N$) gauge theories, even the ones which can not be understood
in
 SU($N$) gauge theories themselves.

In the next section, we give a brief review of the thin and thick center vortex
models. In Sec.III, some general features of the G(2) gauge group are
discussed, and in Sec.IV we apply the improved thick center vortex model, called
the domain structure model, to the G($2$) gauge group. We study the possible reasons
why confinement is observed in the G($2$) gauge group based on its SU(3) subgroups.
Potentials between static sources of higher representations, $14$ and $27$, are
calculated in Sec.V, and Casimir scaling is discussed. We end the paper with a
conclusion.

\section{Thick Center Vortex Model}

The idea of describing confinement using vortices goes back to 't Hooft \cite{Hooft1979}.
In four dimensions vortices are topological field configurations of the vacuum
forming closed surfaces which can be linked to Wilson loops. Wilson loops are
gauge-invariant observables obtained from the holonomy of the gauge connection
around given loops. Confinement is obtained from random fluctuations in the linking
number. A vortex piercing a Wilson loop contributes with a center element $Z$
somewhere between the group elements $U$ of the gauge group
\begin{equation}
\label{Wilson}
 W(C)=\mathrm{Tr}[UUU\ldots U]\longrightarrow \mathrm{Tr}[UUU\ldots(Z)\ldots U].
\end{equation}
 Because center elements commute with all the members of the group, the location of
$Z$ in Eq.(\ref{Wilson}) is arbitrary. The effect of piercings on Wilson
loops can be well understood in the gauge group SU($2$), which has $-I$ as the only
nontrivial center element. $I$ is the $2 \times 2$ unit matrix. Vortices piercing
the loop an even number of times or not piercing it at all, do not have any effect on
$W(C)$. Odd linking numbers change the sign of $W(C)$. One can derive a formula for
the string tension $\sigma$ assuming that vortices are thin and pierce Wilson loops
in single plaquettes with independent probabilities $f$. We get
\begin{equation}
\label{su2-wilson}
\langle W(C)\rangle=\prod \{(1-f)+f(-1)\}\langle W_0(C)\rangle=\exp[-\sigma (c)
A]\langle W_0(C)\rangle
\end{equation}
assuming that $\langle W_0(C)\rangle$ is the expectation value of the loop when no
vortex pierces this loop and the string tension is
\begin{equation}
 \sigma=-\frac{1}{A} \ln(1-2f).
\end{equation}
The vortex model works very well for the fundamental representation, and it gives the
asymptotic string tension correctly. It also works well for the adjoint
representation at large distances, where one expects a screened potential. Since we
have a trivial center element for the adjoint representation, the string tension
must be zero according to Eq.(\ref{Wilson}). In other words, $Z$ is unity and
it does not change the Wilson loop. However, lattice calculations show an
intermediate string tension for higher representations
\cite{Higher,Deld1999,Bali2000}, which can not be predicted by the thin  center
vortex model or Eq.(\ref{Wilson}). M. Faber $\it{et ~al.}$ \cite{Fabe1998}
have improved the model to the thick center vortex model to be able to reproduce the
intermediate string tensions. They have given a thickness to the
't Hooft vortex. Mathematically, they have replaced the center element $Z$ by an
element of the gauge group, $G$,
\begin{equation}
 W(C)=\mathrm{Tr}[UUU\ldots U]\longrightarrow \mathrm{Tr}[UUU\ldots G \ldots U]
\end{equation}
where G is
\begin{equation}
 G(x,s)=\exp(i \alpha_c(x)\overrightarrow{n} \cdot \overrightarrow{L} ).
 \label{G}
\end{equation}
$L_i$'s are the generators of the group in the representation $j$, $\overrightarrow{n}$
is a unit vector, and $S$ is an SU($N$) element of the group in the representation
$j$. $\alpha_c(x)$ gives the profile of the vortex, and it depends on that fraction
of the vortex which is pierced by the loop. It depends on the shape of the loop $C$
and the position of the center of the vortex relative to the perimeter of the loop.

If the Wilson loop is linked by $m$ vortices, centered at positions $x_1, x_2,...,
x_m$, then
\begin{equation}
  W(C)\longrightarrow W[C;\{x_i,S_i\}]=\mathrm{Tr}[UUU\ldots UG\{x_a,S_a\}U\ldots\\
  UG\{x_b,S_b\}U\ldots UG\{x_m,S_m\}U].
\end{equation}
 To obtain the average Wilson loop, $W(C)$ must be averaged over different gauge
group orientations $S_i$. For simplicity, we first look at the SU($2$) case.
Equation (\ref{G}) for SU($2$) is
\begin{equation}
 G(x,s)=\exp(i \alpha_c(x)\overrightarrow{n} \cdot \overrightarrow{L} )= S \exp(i
\alpha_{c}(x) L_3)S^\dagger.
 \label{su2-G}
\end{equation}
$L_{3}$ is the element of the Cartan subalgebra.Independently averaging every
$G\{x_a,S_a\}$ over orientations in the group manifold specified by $S_a$, we get
\begin{equation}
\bar{G}(a)=\int \!  S\exp[i\alpha L_3]S^\dagger\, \mathrm{d} S= g_j[\alpha_c] I_{2j+1}
\end{equation}
where
\begin{equation}
g_j[\alpha_c]=\frac{1}{2j+1}\mathrm{Tr}(\exp[i\alpha L_3]).
\end{equation}
Thus, averaging the Wilson loop over all $S_a$ leads to \cite{Fabe1998}
\begin{eqnarray}
\label{Wils}
 \langle W(C)\rangle=\prod_x \{(1-f)+fg_j[\alpha_c]\}\langle W_0(C)\rangle  \\
 \label{wils-su2}
  =\exp[\sum_x \ln\{(1-f)+fg_j[\alpha_c]\}]\langle W_0(C)\rangle \\
 =\exp[-\sigma_c A]\langle W_0(C)\rangle
\end{eqnarray}
where $\sigma_c$ is
\begin{equation}
\sigma_c=\frac{-1}{A} \sum_x \ln\{(1-f)+fg_j[\alpha_c]\}.
\label{sigma-c}
\end{equation}
For large loops, the center element is located completely inside the loop; thus
$g_j[\alpha_c]=-1$, and the string tension is
\begin{equation}
 \sigma_c=-\frac{1}{A} \ln(1-f-f)=-\ln(1-2f).
\end{equation}
This string tension is the same for all half integer representations $j$. For
integer representations for which a flat potential is expected  at large distances,
the center element is trivial and $g_j[\alpha_c]=1$; thus $\sigma_c=0$
according to Eq.(\ref{sigma-c}).

At intermediate distances where the loop is not large enough to contain the whole
vortex, $\sigma_c$ is \cite{Fabe1998}
\begin{equation}
\sigma_c=\frac{f}{6}\bar{\alpha}_c^2 j(j+1)
\end{equation}
where
\begin{equation}
\bar{\alpha}_c^2=\frac{1}{A}\sum_{x\in A'}\alpha_c^2(x).
\end{equation}
Thus, for small $f$ and small loops, the string tension is proportional to the
eigenvalue of the second order Casimir operator.

Equation (\ref{Wils}) can be generalized to the SU($N$) gauge theory
\begin{equation}
\label{N-Wils}
\langle W(C)\rangle=\prod_x {(1-\sum_{n=1}^N f_n(1-g_r [\vec{\alpha}_c^n (x)]))}
\langle W_0(C)\rangle
\end{equation}
with
\begin{equation}
\label{gr1}
g_r[\vec{\alpha}]=\frac{1}{d_r}\mathrm{Tr}(\exp[i\vec{\alpha} \cdot \vec{H}]).
\end{equation}
$x$ is the location of the center of the vortex, $d_r$ the dimension of the
representation $r$, $f_{n}$ the
probability that any given unit is pierced by a vortex of type $n$, and
$\{H_{i},i=1,2,...,N-1\}$  are the generators spanning the Cartan subalgebra.
$\vec{\alpha}_c^n(x)$ shows the flux profile for the vortex of type $n$. We have
$(N-1)$ types of vortices for any  SU($N$) gauge theory. The potential and the
string tension are given by \cite{Fabe1998}
\begin{equation}
\label{potential}
V(R) = \sum_{x}\ln\left\{ 1 - \sum^{N-1}_{n=1} f_{n}
(1 - {\mathrm {Re}} g_{r} [\vec{\alpha}^n_{C}(x)])\right\}
\end{equation}
and
\begin{equation}
\label{sig}
\sigma_C=-\frac{1}{A}\sum_{x} \ln[{1-\sum_{n=1}^{N-1}f_n(1-{\mathrm {Re}}
g_r[\vec{\alpha}_C^n(x)])}].
\end{equation}

To reproduce the potential or the string tension, one has to find the Cartan
subalgebra and describe the profile function $\vec{\alpha}_C^n(x)$. The flux
profile should have the following properties: It should go to zero for a
 Wilson loop of length $R$ as $x\rightarrow \pm \infty$; if the center of a vortex
is completely inside the loop, it will lead to a maximal multiplicative factor
$\exp({\frac{2\pi i n}{N}})\in Z_{n}$  $(n=1,2,...,N-1)$ corresponding to the
center elements of the group;
 and finally as $ R \rightarrow 0 $ the contribution of the vortex to the loop also
goes to zero , so $\alpha_R(x)\rightarrow 0$.

Various vortex profiles $\vec{\alpha}_C^n(x)$ may be chosen. M. Faber $\it{et~ al.}$
used, for the SU($2$) case,
\begin{equation}
\vec{\alpha}_C^n(x)=\vec{N}^n[1-\tanh(ay(x)+\frac{b}{R})]
\end{equation}
where $n$ indicates the vortex type, $a , b$ are arbitrary constants, and $y(x)$ is
\begin{equation}
y(x)=\begin{cases} -x & \lvert R-x\lvert>x \\x-R & \lvert R-x\lvert\leqslant x
\end{cases}
\end{equation}
$y(x)$ is the nearest distance of $x$ from the timelike side of the loop. The
normalization constant $\vec{N}^n$ is obtained from the maximum flux condition, where
the loop contains the vortex completely,

\begin{equation}
\label{normal}
 \exp(i\vec{\alpha}^n\cdot\vec{H})=z_n I
\end{equation}
with
\begin{equation}
z_n=e^{\frac{2\pi in}{N}} \in Z_N
\end{equation}
and $I$ is the unit matrix.

The thick center vortex model has worked well for the SU($2$), SU($3$), and SU($4$)
gauge theories \cite{Fabe1998}, \cite{Deld}. A linear potential in  rough agreement
with Casimir scaling is observed for all representations, and the asymptotic string
tension  proportional to the $N$-ality of the given representation has been
reported. To increase the length of the linear regime at intermediate distances, J.
Greensite $\it{et~ al.}$ \cite{Green2007} have suggested  postulating a kind of
domain structure in the vacuum, with magnetic flux in each domain quantized in units
corresponding to the elements of the center subgroup, including the identity
element. They have applied this idea to the SU($2$) gauge group which has one
trivial and one nontrivial center element. Thus, instead of one vortex, they use
two kinds of domain structures corresponding to the center elements, the trivial and
nontrivial elements. To obtain the Wilson loop, Eq. (\ref{Wils}) should be
changed to
\begin{equation}
\langle W(C)\rangle=\prod_x \{(1-f_1-f_0)+f_0g_j[\alpha_C^0]
+f_1g_j[\alpha_C^1]\}\langle W_0(C)\rangle .
\end{equation}
$f_1$ and $f_0$ are the probabilities of plaquettes that belong to domains
corresponding to the nontrivial and trivial center elements, respectively.
$\alpha_C^0$ and $\alpha_C^1$ define the profile function for domain structures
associated with the trivial and nontrivial center elements. To apply the domain
structure model to the SU($N$) gauge group, one has to change $n$ in Eqs.
(\ref{N-Wils}), (\ref{potential}) and (\ref{sig}) from zero to the number of
nontrivial center elements, where zero corresponds to the trivial center element.

G($2$) lattice gauge calculations show confinement for intermediate distances
\cite{Olejnik2008}. The center vortex model can predict correctly the screening of
the potentials at large distances since the center of the G($2$) gauge group is
trivial. But, what accounts for the intermediate string tensions? We apply the
domain structure model, which includes the trivial center element, to the G($2$)
gauge group in Sec.IV and discuss the possible reasons for the linear
potential at large distances. But first, in Sec.III, we present some general
features of the gauge group G($2$).

\section{General properties of G($2$) gauge theories}

G($2$) is one of the exceptional Lie groups which is its own covering group
\cite{Pepe}.
The rank of G($2$) is 2 and it has 14 generators,two of which are
simultaneously diagonal.
Its fundamental representation is seven dimensional, and the dimension of the  adjoint
representation is 14. The group G($2$) is real, and it is a subgroup of SO($7$)  with
rank 3 and 21 generators.
The determinant of the $7 \times 7$ real orthogonal matrices $U$ of the group
SO($7$) is $1$, and
\begin{equation}
UU^{\dagger}=1.
\end{equation}
G($2$) elements satisfy a constraint called the
cubic constraint,
\begin{equation}
T_{abc}=T_{def}U_{da}U_{eb}U_{fc}
\end{equation}
$T$ is a totally antisymmetric tensor, and its nonzero elements are
\begin{equation}
T_{127} = T_{154} = T_{163} = T_{235} = T_{264} = T_{374} = T_{576} = 1.
\end{equation}
Because of these constraints the number of generators of G($2$) is reduced to $14$.

A quark in the fundamental representation $\{7\}$ of G($2$) can be color screened by
gluons \cite{Pepe},
\begin{equation}
\label{product7}
\{7\} \otimes \{14\} \otimes \{14\} \otimes \{14\}=\{1\} \oplus ...
\end{equation}
Therefore, in G($2$) gauge theories, the string tension between static quarks can be
broken by gluons without the presence of dynamical quarks, as is necessary in SU($N$)
gauge theories.

SU($3$) is a subgroup of G($2$). Under SU(3) subgroup transformations, the seven and
$14$-dimensional
representations  of G($2$) decompose into the SU($3$) fundamental and adjoint
representations
\begin{equation}
\label{7t03}
\{7\} = \{3\} \oplus \{\overline 3\} \oplus \{1\} ,
\end{equation}
\begin{equation}
\label{14to3}
\{14\} = \{8\} \oplus \{3\} \oplus \{\overline 3\}.
\end{equation}

The second equation may be interpreted in such a way that the $14$ gluons of G($2$)
consist of the usual eight gluons of SU($3$) plus six additional gluons which
transform like the SU($3$) fundamental quark and antiquark. One of the differences
between these six gluons and the SU($3$) quarks is that the former are bosons while
the latter are fermions. Because all G($2$) representations are real, the
fundamental representation is equivalent to its complex conjugate; therefore quarks
and antiquarks are identical.

Simultaneous diagonal generators of this representation are \cite{Pepe}
\begin{eqnarray}
\label{t3}
H^3=\frac{1}{\sqrt{8}}(P_{11}-P_{22}-P_{55}+P_{66}),\\
\label{t8}
H^8=\frac{1}{\sqrt{24}}(P_{11}+P_{22}-2P_{33}-P_{55}-P_{66}+2P_{77})
\end{eqnarray}
where $ (P_{ij})_{\alpha \beta} =\delta_{i \alpha} \delta_{j \beta} $ , and $ \alpha
, \beta$ indicate the row and the column of the matrices, respectively.
From Eqs. (\ref{t3}) and (\ref{t8}), it can be shown that these generators have
SU($3$) Cartan generators, and they can be written as
\begin{equation}
\label{cartan}
H^a = \frac{1}{\sqrt{2}} \left( \begin{array}{ccc} \lambda^a &0 & 0 \\
0 & \; 0 & 0\\ 0 & 0 & -(\lambda^a)^*  \
\end{array} \right).
\end{equation}
$\lambda^a$ ($a=3,8$) are the two diagonal SU(3) generators.
Since SU($3$) is a subgroup of the G($2$) gauge group, the G($2$) weight diagram can
be obtained by a superposition of $3$, $1$, and $\bar{3}$. Therefore, in the SU($3$)
subgroup of G($2$), the center elements of G($2$) can be constructed from Z($3$),
the center of SU($3$)
\begin{equation}
\label{Z}
Z= \left(\begin{array}{ccc} zI_{3\times3} & 0 & 0 \\0 &1 &0 \\ 0 & 0 & z^*I_{3\times3}
\end{array} \right).
\end{equation}
$I_{3\times3} $ is a $ 3\times3 $ unit matrix and $z\in\{1,e^{\pm\frac{2\pi i}{3}}\}
$ are elements of $Z(3)$.
The three Z matrices of Eq.(\ref{Z}) commute with the eight generators of the
SU($3$) subgroup of G($2$) but not with the remaining six generators. This is why
the centers Z given by Eq. (\ref{Z}) are not the center elements of G($2$). The
center elements should commute with all the generators. Therefore, the center of
G($2$) should be trivial and contains only the identity. This is true for higher
representations of G($2$) as well. Since the center element of G($2$) is trivial
for all representations, at large distances the potentials between objects in all
representations are flat and screening happens. Equation (\ref{product7}) shows
this fact for the seven dimensional representation. Another consequence is that the
concept of $N$-ality does not apply to the G($2$) gauge group, in contrast to the
SU($N$) gauge groups.

In the next section, we calculate the potential between two G($2$) quarks using the
domain structure model.

\section{G($2$) and vacuum domain structures}

Numerical calculations \cite{Olejnik2008} show a linear behavior at intermediate
distances and a flat potential at large distances for the potential between two
quarks in the fundamental and higher representations of the G($2$) gauge theory.
Even though G($2$) does not have any nontrivial center element, its trivial center
element can explain the flat potential according to the thick center vortex model.
Based on this model no vortex means no string tension at large distances, and this is
in agreement with the flat potential for the G($2$) quarks at large distances. In
this section we reproduce the potential in a G($2$) quark antiquark
pair using the idea of domain structures of the vacuum in the thick center vortex model.

Greensite $\it{et~ al.}$ have introduced domain structures to increase the length of
the linear regime in the thick center vortex model \cite{Green2007}. They have
suggested that the domain structure idea  can be applied to the G($2$) gauge group
which has only a trivial center.
With this motivation, we apply the idea to calculate the G($2$) potentials
explicitly. In SU($N$), zero $N$-ality representations have zero string tensions at
large distances, while they have linear potentials at intermediate distances. The
thick center vortex model has been able to reproduce this behavior, and we have been
motivated to use this model to reproduce G($2$) potentials which show the same
behavior according to the lattice results but with the idea of domain structures
instead of center vortices. Thickening the vortices leads to the confinement for
higher SU($N$) representations. Since G($2$) has only a trivial center element, we
apply the thick center vortex model to this group with the constraint that the
total magnetic flux measured by the Wilson loop holonomy is quantized in terms of
its trivial center element.

In the domain structure model, the vacuum of the Yang-Mills theory is filled with
field configurations called domain structures, corresponding to trivial and
nontrivial center elements. For example, for SU($2$)there exist two type of
domains, one associated with  the trivial one, $I$, and one associated
with the nontrivial one, $-I$. The main difference between the trivial domain and
other domains is the lack of a Dirac-3 volume for trivial domains.

\begin{figure}[b]
\begin{center}
\vspace{70pt}
\resizebox{0.57\textwidth}{!}{
\includegraphics{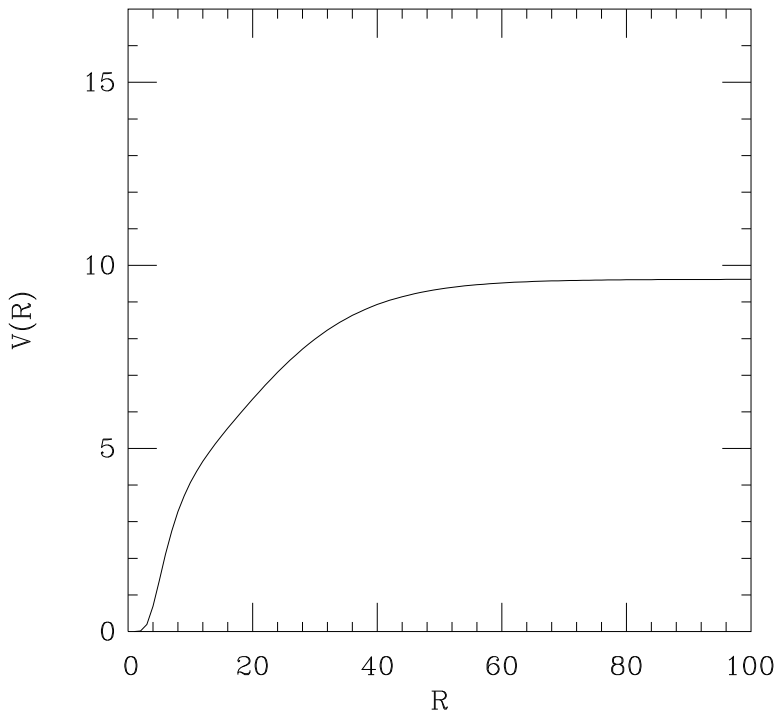}}
\vspace{-20pt}
\caption{\label{G2pot}
Potentials between two G($2$) static quarks obtained from the domain structure
model. The potential is screened at large distances, in agreement with the fact that
the gluons of the G($2$) gauge group can couple to the quarks in the fundamental
representation and give a singlet. A linear potential is observed for the
intermediate distances.}
\end{center}
\end{figure}

Equation (\ref{potential}) can be used in the SU($N$) gauge group for the modified
thick center vortex model, if $n$ starts from zero instead of 1, where $n=0$
corresponds to the trivial center element. For the G($2$) case with only a trivial
center element and no other nontrivial center elements, $V(R)$ is given by
\begin{equation}
\label{G2potential}
V(R) = \sum_{x}\ln\left\{ 1 - f_{0}(1 - {\mathrm {Re}} g_{r}
[\vec{\alpha}^0_{C}(x)])\right\}.
\end{equation}
where $f_{0}$ is the probability that any given unit is pierced by a trivial vacuum
domain. $g_{r}$ has the same form as in Eq. (\ref{gr1}),
\begin{equation}
\label{gr2}
g_r[\vec{\alpha}]=\frac{1}{d_r}\mathrm{Tr} (\exp[i\vec{\alpha} \cdot \vec{H}]).
\end{equation}
We use the flux profile
\begin{equation}
\alpha_i^0(x)=N_i^0[1-\tanh(ay(x)+\frac{b}{R})].
\label{alpha0}
\end{equation}
The zero index indicates the trivial domain of G($2$), and the index $i$ is the index
of the Cartan generators. $a$ and $b$ are free parameters of the model.
The normalization factor is obtained by using the maximum flux condition of Eq.
(\ref{normal}). For G($2$), the equation is changed to
\begin{equation}
 \exp(i\vec{\alpha}^n \cdot \vec{H})= I.
\label{expalpha}
\end{equation}
Therefore,
\begin{equation}
 \exp(\alpha_1^{max}H_1+\alpha_2^{max}H_2)=I=\exp(2\pi i).
\end{equation}
Using the Cartan subalgebra of G($2$) from Eq. (\ref{cartan}), $\alpha_1^{max}
$ and $\alpha_2^{max} $ are obtained,
\begin{equation}
\begin{array}{c}
2\pi=\frac{\alpha_1^{max}}{\sqrt{8}}+\frac{\alpha_2^{max}}{\sqrt{24}}
\\2\pi=\frac{-\alpha_1^{max}}{\sqrt{8}}+\frac{\alpha_2^{max}}{\sqrt{24}}
\end{array}\}
\Longrightarrow ~~\alpha_1^{max}=0, ~~\alpha_2^{max}=2\pi \sqrt{24}.
\end{equation}
Thus, $\alpha_1^0(x)$ and $\alpha_2^0(x)$ are
\begin{equation}
 \alpha_1^0(x)=0,
\end{equation}
\begin{equation}
\alpha_2^0(x)=\frac{\pi}{\sqrt{24}}[1-\tanh(ay(x)+\frac{b}{R})].
\end{equation}
Putting everything together, the potential between two G($2$) static quarks is
obtained from Eqs. (\ref{G2potential}) to (\ref{alpha0}) and is plotted in
Fig. (\ref{G2pot}). The free parameters $a$, $b$, and $f$ are chosen to be $0.05$,
$4$, and $0.1$, respectively. The potential is screened at large distances as
expected from the model. At large distances the domains are contained completely
inside the loops, and the corresponding contribution, which is the trivial center
element, does not change the Wilson loop; therefore, the string tension is zero.
At intermediate distances a linear regime is observed. For this regime, some
fraction of the trivial domain is located inside the loop. Therefore it can make
a nontrivial contribution to the Wilson loop and lead to a nonzero string
tension.
\begin{figure}[b]
\begin{center}
\vspace{70pt}
\resizebox{0.57\textwidth}{!}{
\includegraphics{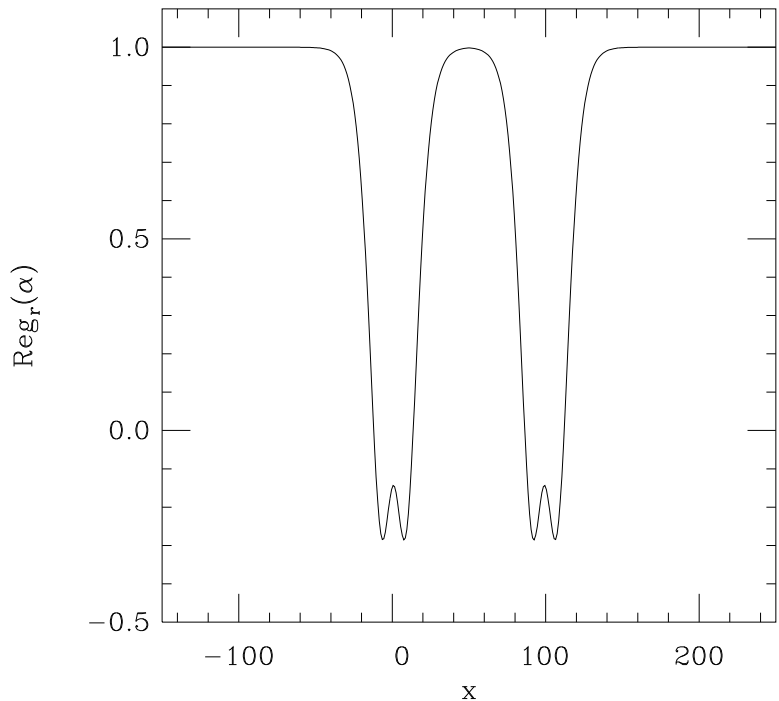}}
\vspace{-20pt}
\caption{\label{gr-fig}
${\mathrm {Re}}(g_{r}[\alpha^0_{2}(x)])$ versus $x$ is plotted for the fundamental
representation of the G($2$) gauge group. It varies between $1$ and $-0.28$. The
value of $1$ indicates the situation where the Wilson loop either does not link the
vortex or it links the vortex completely. ${\mathrm {Re}}(g_{r}
[\alpha^0_{2}(x)])=-0.28$ can be explained by the SU($3$) subgroup of G($2$) as
explained in the text. }
\end{center}
\end{figure}
\begin{figure}[b]
\begin{center}
\vspace{70pt}
\resizebox{0.57\textwidth}{!}{
\includegraphics{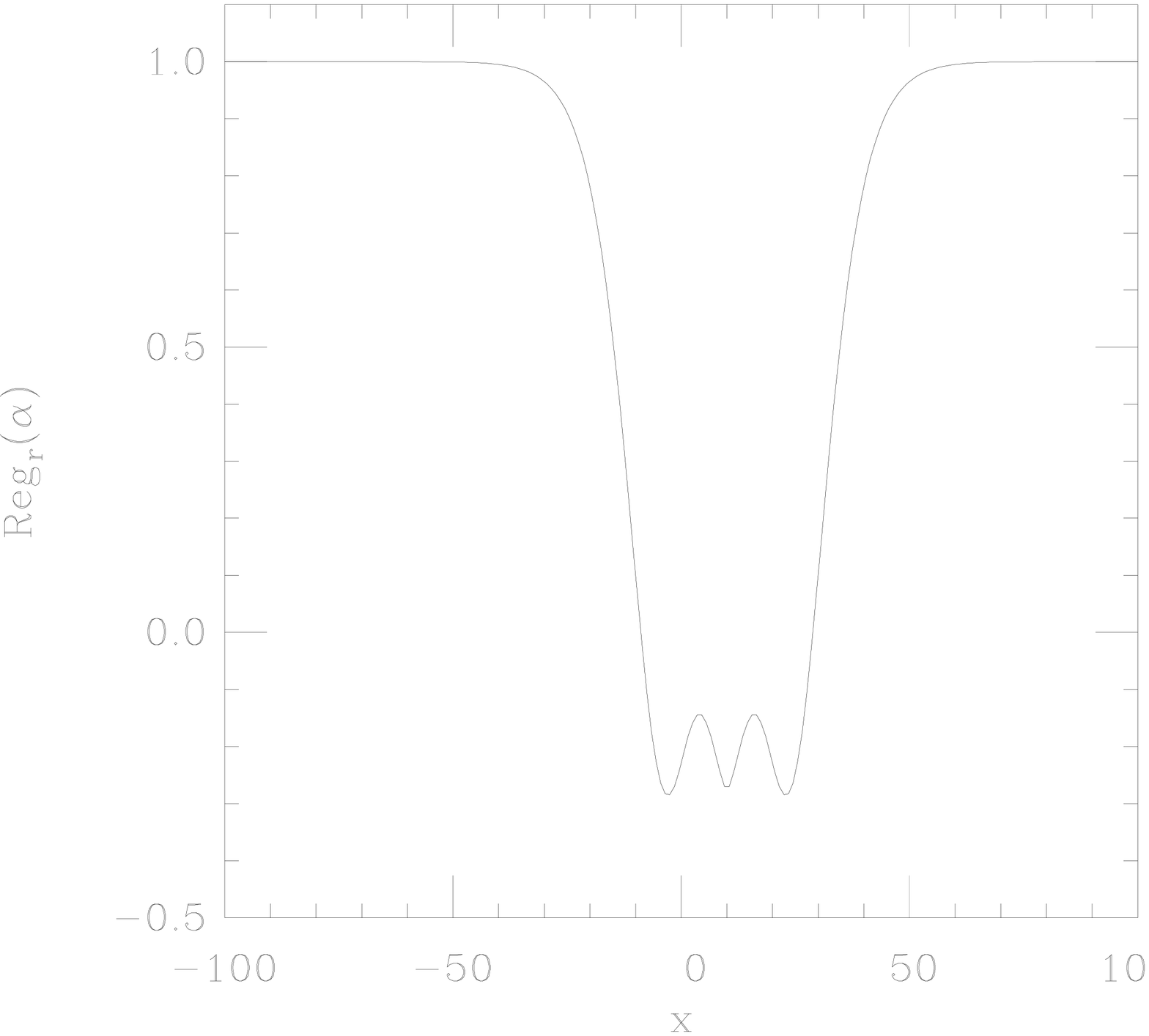}}
\vspace{-20pt}
\caption{\label{gr-fig2}
The same as Fig. (\ref{gr-fig}) but for $R=20$. The minimum of ${\mathrm
{Re}}(g_{r} [\alpha^0_{2}(x)])$ reaches $-0.28$ for this intermediate $R$ as well.}
\end{center}
\end{figure}

To study the possible reasons for the linear part, we plot ${\mathrm {Re}}(g_{r}
[\alpha^0_{2}(x)])$ versus $x$ for the Wilson loop, with various spatial
dimensions of $R$. Figure (\ref{gr-fig})  plots ${\mathrm {Re}}(g_{r}
[\alpha_2^0(x)])$ versus $x$ for $R=100$. $x$ shows the location of the center of
the vortex. The left leg of the Wilson loop is located at zero, and the right leg is
located at $x=R$.  The maximum of ${\mathrm {Re}}(g_{r})$ is equal to 1
and indicates either the situation where the domain does not link to the Wilson
loop or the one where it is located completely inside the loop. For the profile we are using, the
size of the domain is proportional to the inverse of the parameter $a$, and it is
about $20$. Therefore, when the Wilson loop spatial length is equal to $100$, we are
sure that a complete link is established between the Wilson loop and the vortex.
This fact is observed from the figure, since ${\mathrm {Re}}(g_{r})$ reaches $1$
when the center of the vortex is located at $x\approx50$.

${\mathrm {Re}}(g_{r})$ reaches a value of 1 when the vortex is completely inside a
Wilson loop. In SU($N$) the result would be different, and a non-trivial center
element is obtained. Another interesting feature of the figure is the minimum of
${\mathrm {Re}}(g_{r})$ which is around $-0.28$. In general, for
 the SU($N$) cases ${\mathrm {Re}}(g_{r})$ varies from $1$, corresponding to the
trivial center element, to the values corresponding to the nontrivial center
elements. Since G($2$) does not have any nontrivial center element, predicting the
amount of the lower limit is not clear.
The minimum of ${\mathrm {Re}}(g_{r})$, can be interpreted using the SU(3) subgroup
of G($2$).
SU($3$) has three center elements (domains): $e^{\frac{n\pi i}{3}}$,  where $n$
changes from zero to 2. ${\mathrm {Re}}(g_{r} [\alpha^0_{2}(x)])$, corresponding to
these center elements, varies between $1$ and $-0.5$ since
\begin{equation}
 e^{\frac{2\pi i}{3}}=-0.5+\frac{\sqrt{3}}{2}i.
\end{equation}
Now, calculating ${\mathrm {Re}}(g_{r} [\alpha^0_{2}(x)])$ of the G($2$) gauge group
using its SU($3$) subgroup and normalizing it with the dimension of the subgroup, we
get the minimum of ${\mathrm {Re}}(g_{r}[\alpha^0_{2}(x)])$,
\begin{equation}
 {\mathrm {Re}}g_{r}
([\alpha^0_{2}(x)])_\mathrm{min}=\frac{1}{7}(\mathrm{Tr}(e^{i\alpha \cdot
H})_\mathrm{min})= \frac{1}{7}\mathrm{Tr} \left ( \begin{array}{ccc}
{\acute{g_R}(\alpha)_\mathrm{min}\times d_r} & 0 & 0 \\ 0 & 1 & 0
\\0 & 0 & {\acute{g_R}(\alpha)_\mathrm{min}\times d_r}
\end{array} \right ).
\end{equation}
where ${\acute{g_R}}(\alpha)_\mathrm{min}$ is the minimum of ${\mathrm{Re}}(g_{r})$
for the SU($3$) subgroup, and it is equal to $-0.5$ as mentioned above. $ d_r $ is
the dimension of the subalgebra which is equal to 3. Thus, the minimum of
${\mathrm{Re}}(g_{r})$ for the group G($2$) is obtained as
\begin{equation}
  {\mathrm{Re}}(g_{r})_\mathrm{min}=\frac{1}{7}(\mathrm{Tr}(e^{i\alpha \cdot
H})_\mathrm{min})=\frac{1}{7}(-1.5+1-1.5)=-0.28.
\end{equation}
Some other local minimums are observed for ${\mathrm{Re}}(g_{r})$ in Fig.
(\ref{gr-fig})  which may be interpreted by the SU($2$) subgroups of G($2$).
Reaching the minimum of $-0.28$ happens for the intermediate distances $R$ as well.
Figure (\ref{gr-fig2})  shows ${\mathrm {Re}}(g_{r} [\alpha^0_{2}(x)])$ for $R=20$.

Going back to Fig.(\ref{G2pot}), because of the trivial center of the G($2$) gauge
group, at large distances a screened potential is expected from thin or thick center
vortex models. However, the linear regime is observed as a result of using thick
vortices, as seen for the SU($N$) higher representations with zero N-ality. On
the other hand, studying the flux profile, ${\mathrm {Re}}(g_{r}
[\alpha^0_{2}(x)])$, the role of the SU($3$) subgroup of G($2$) must be considered
as another possible reason to interpret the linear behavior. One may claim that,
at intermediate distances, the SU($3$) subgroup of G($2$) has a dominant role, and
this fact leads to the observation of a linear potential at this regime.

Pepe $\it{et~ al.}$ \cite{Pepe} have studied the role of the
SU($3$) subgroup  with a Higgs mechanism. They have exploited a scalar Higgs field in
representation $7$ of
G($2$), which breaks G($2$) to SU($3$). This allows for interpolation between theories
with exceptional confinement like G($2$) and theories with ordinary confinement like
SU($3$). An interpretation of G($2$) confinement via its SU($3$) subgroup and using
the domain structure model needs more investigation, and the present paper may be
considered as one of the starting points.

In the next section, we find the potentials between heavy sources for the higher
representations $14$ and $27$, and we discuss Casimir scaling of the string tension.

\section{Potentials between static G($2$) sources of higher representations}

\begin{figure}[]
\begin{center}
\vspace{70pt}
\resizebox{0.87\textwidth}{!}{
\includegraphics{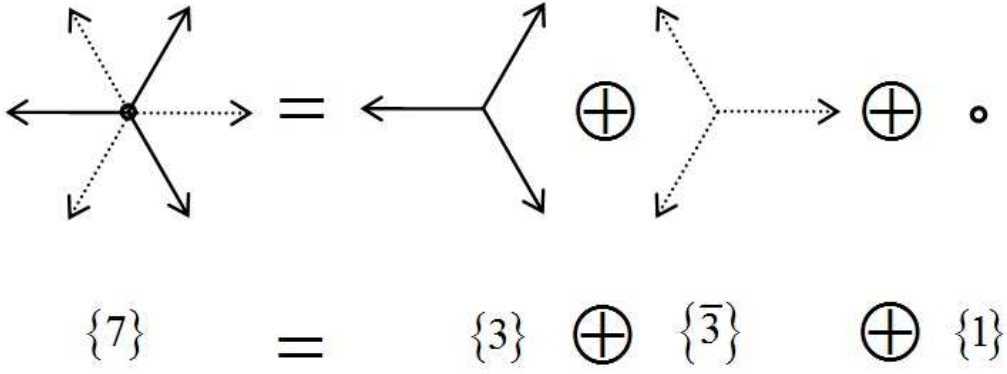}}
\caption{\label{weight7}Weight diagram of the fundamental representation of G($2$)
decomposed into its SU($3$) subgroups.
 }
\end{center}
\end{figure}
Sources with dimensions $14$ and $27$ are obtained by
\begin{equation}
\{7\} \otimes \{7\} =\{1\} \oplus \{7\} \oplus \{14\} \oplus \{27\}.
\label{rep27}
\end{equation}
To calculate the potentials with the domain structure model, one has to use Eqs.
(\ref{G2potential}) and (\ref{gr2}). As Eq. (\ref{gr2}) demands, $H$, the
Cartan subalgebra of G($2$) must be calculated for the $14$- and $27$- dimensional
representations. We use the weight diagrams of the SU($3$) gauge group, and we
decompose each representation of the  G($2$) gauge group to its SU($3$) subgroups to
obtain the weight diagrams for G($2$) representations. It is clear from Eq.
(\ref{cartan}) that the weight diagram of the G($2$) fundamental representation can be
obtained by Fig. (\ref{weight7}). This is because the fundamental representation
of G($2$) can be decomposed into its SU($3$) subgroup representations
\begin{equation}
\{7\}= \{3\} \oplus \{\bar{3}\} \oplus \{1\}
\end{equation}
where $3$ and $\bar{3}$ show the fundamental representations of the SU($3$) group.

\begin{figure}[]
\begin{center}
\vspace{70pt}
\resizebox{0.77\textwidth}{!}{
\includegraphics{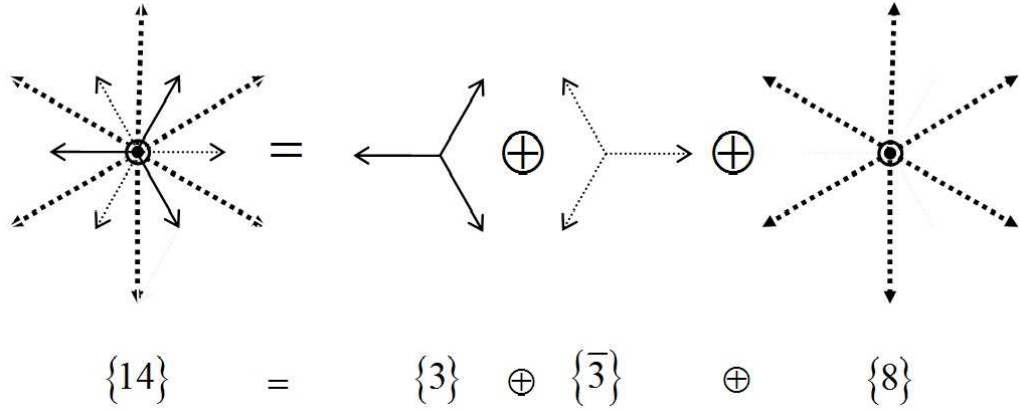}}
\caption{\label{weight14}Weight diagram of the adjoint ($14$-dimensional)
representation of G($2$) decomposed into its SU($3$) subgroups.
 }
\end{center}
\end{figure}

For the adjoint representation of G($2$) with dimension $14$, the decomposition is
as follows:
\begin{equation}
\{14\} = \{3\} \oplus \{\bar{3}\} \oplus \{8\}.
\label{3comb}
\end{equation}
The $8$ indicates the adjoint representation of SU($3$). Thus, the Cartan subalgebra of
the adjoint representation can be calculated from the weight diagrams of Fig.
(\ref{weight14}) and Eq. (\ref{3comb}),
\begin{equation}
\label{cartan14}
H_{14} = \frac{1}{\sqrt{8}} \left( \begin{array}{ccc} \lambda_3 &0 & 0 \\
0 & \; -{\lambda_{\bar{3}}}^* & 0\\ 0 & 0 & \lambda_8 \
\end{array} \right).
\end{equation}
where $\lambda_3$, $\lambda_{\bar{3}}$, and $\lambda_8$ are the Cartan subgroups of
SU($3$) for the representations $3$, ${\bar{3}}$, and $8$, respectively.
\begin{figure}[b]
\begin{center}
\vspace{70pt}
\resizebox{0.97\textwidth}{!}{
\includegraphics{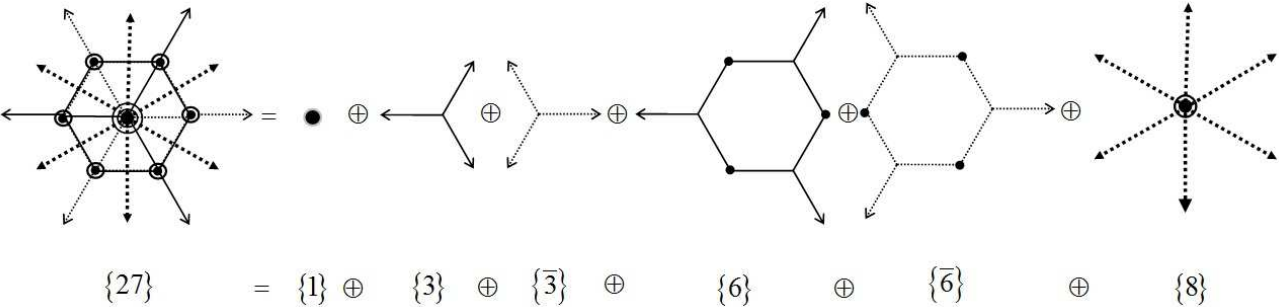}}
\vspace{-20pt}
\caption{\label{weight27}Weight diagram of the representation $27$ of G($2$)
decomposed into its SU($3$) subgroups.
 }
\end{center}
\end{figure}

Representation $27$ of the G($2$) gauge group is obtained from Eq. (\ref{rep27}).
But like the fundamental and adjoint representations of G($2$), the weight diagram
of representation $27$ can be obtained by decomposition to its  SU($3$)
subgroups. Figure (\ref{weight27})  shows this decomposition, where
$6$ and $\bar{6}$ are the six dimensional representations of SU($3$).
In fact,
\begin{equation}
\{27\}=\{8\} \oplus \{6\} \oplus \{\bar{6}\}    \oplus \{3\} \oplus \{\bar{3}\}
\oplus \{1\}.
\label{27dec}
\end{equation}
Therefore, the Cartan subalgebra of representation $27$ is
\begin{equation}
\label{cartan27}
H_{27} = \frac{1}{\sqrt{18}} \left( \begin{array}{cccccc} \lambda_8 &0 & 0  &0 & 0
&0  \\
0 &  \lambda_6 & 0 & 0  &0 & 0 \\ 0 & 0 & -{\lambda_{\bar{6}}}^* & 0  &0 & 0 \\ 0 &
0 & 0 &
 \lambda_3 & 0 &0 \\ 0 & 0 &  0 & 0 &-{\lambda_{\bar{3}}}^*  & 0   \\ 0 & 0 & 0 &  0
& 0 & 0 \
\end{array} \right).
\end{equation}
We recall that for all representations of G($2$) we use the normalization condition
\begin{equation}
\label{normaliz}
\mathrm{Tr}[T_{a}T_{b}]=\frac{1}{2}\delta_{ab}.
\end{equation}
\begin{figure}[t]
\begin{center}
\vspace{70pt}
\resizebox{0.57\textwidth}{!}{
\includegraphics{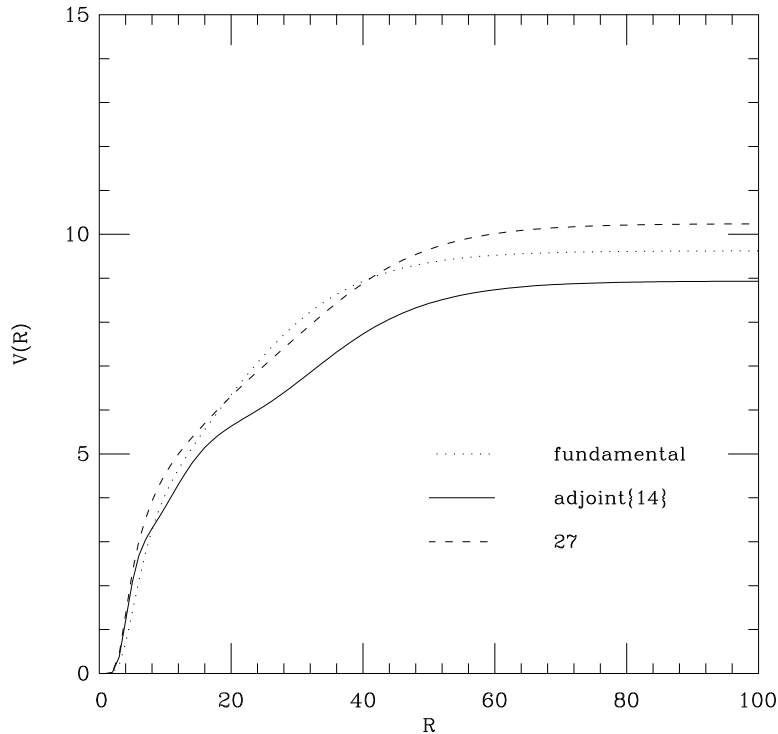}}
\vspace{-20pt}
\caption{\label{all}
Potentials between two G($2$) static sources in the fundamental, $14$ and $27$
dimensional representations. The potentials are screened at large distances in
agreement with the fact that gluons of the G($2$) gauge group can couple to the
initial sources and a singlet is produced. For each representation, a linear
potential is observed at intermediate distances.}
\end{center}
\end{figure}
\begin{figure}[]
\begin{center}
\vspace{70pt}
\resizebox{0.57\textwidth}{!}{
\includegraphics{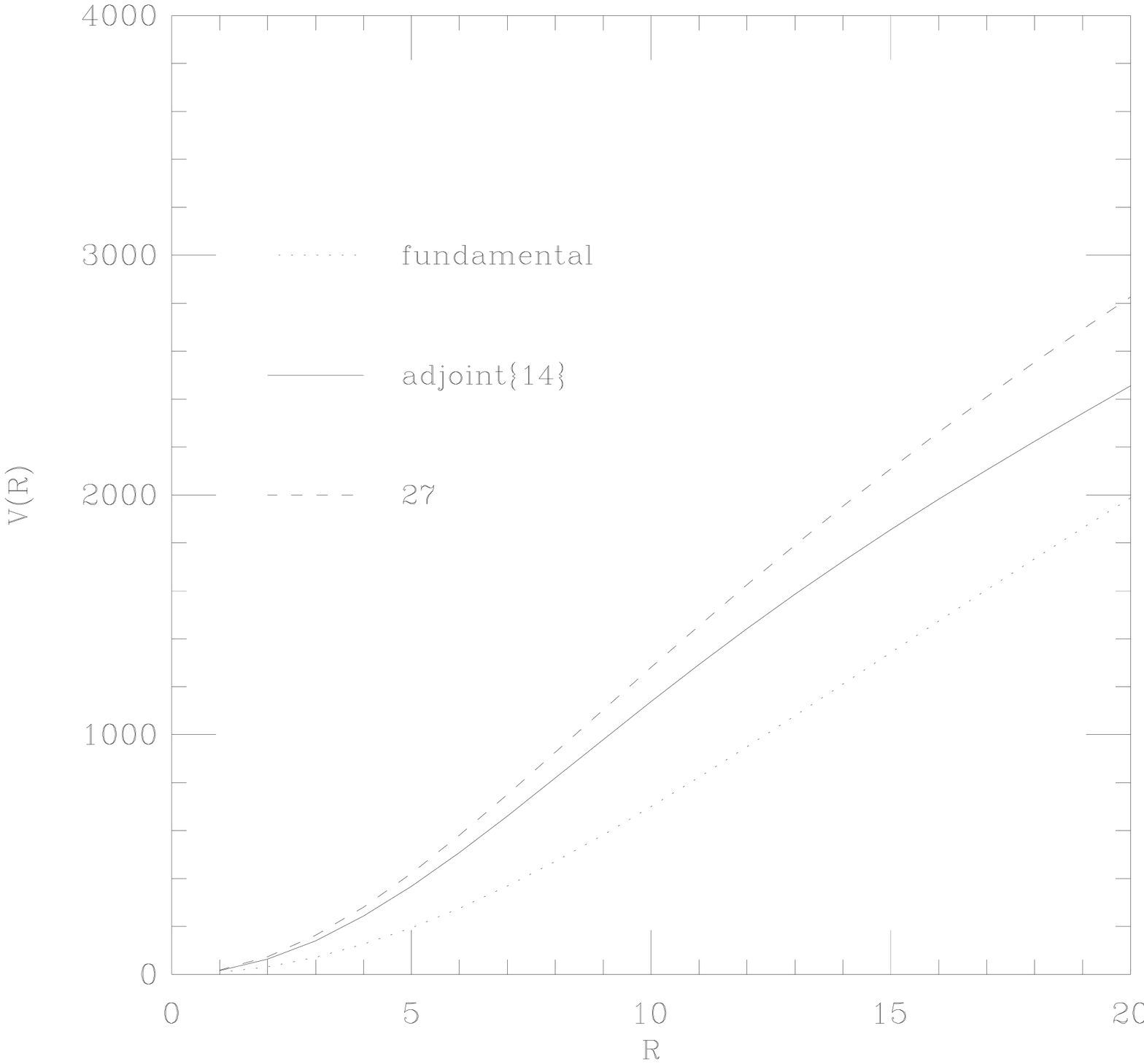}}
\vspace{-20pt}
\caption{\label{short-dist}
Potentials between two G($2$) static sources in the fundamental, $14$ and $27$
dimensional representations at intermediate distances. The ratio of the string
tension of each representation to the string tension of the fundamental
representation is qualitatively in rough agreement with the Casimir ratio.}
\end{center}
\end{figure}
The next step is to find the normalization factors in Eq.
(\ref{alpha0}) for these two new representations. We use Eq.
(\ref{expalpha}), with $\vec{H}$ the Cartan subgroups of representations $14$
and $27$. For both representations $\alpha_{1}^{max}=0$ and $\alpha_{2}^{max}$
is equal to $\pi\sqrt{24}$ for the adjoint and $6\pi\sqrt{24}$ for the $27$
representation
.
Now, we are ready to calculate the potentials from Eq. (\ref{G2potential}) with
the profile function of Eq. (\ref{alpha0}). Fig. (\ref{all}) plots the
potentials for the fundamental, adjoint, and $27$ representations. The free
parameters $a$, $b$, and $f$ are chosen to be $0.05$, $4$, and $0.1$, respectively.
The potentials are screened at large distances, as expected from the fact that G($2$)
has only a trivial center. In fact, the representations $14$ and $27$ lead to a
singlet after they combine with gluons that popped out of the vacuum. This happens when
the distance between static sources is large enough and the potential between
sources is big enough to create a pair of heavy gluons.
\begin{equation}
\{7\} \otimes \{14\} \otimes \{14\} \otimes \{14\}=\{1\} \oplus ...
\label{7singlet}
\end{equation}
\begin{equation}
\{14\} \otimes \{14\} =\{1\} \oplus ...
\label{14singlet}
\end{equation}
\begin{equation}
\{27\} \otimes \{14\} =\{7\} \oplus \{14\} \oplus \{27\} \oplus ...
\label{27singlet}
\end{equation}
One of the interesting points of the figure is that the sources in the fundamental
representation are screened at higher energy than the adjoint sources. This can be
explained by the above equations which indicate that a fundamental quark interacting
with three gluons leads to a singlet, while adjoint sources are screened by
interacting with one gluon. Therefore, screening happens at higher energies for the
fundamental representation compared to the adjoint representation. The sources of
the $27$-dimensional representation do not give a singlet after combining with a
pair of gluons. Screening happens if the adjoint sources produced from Eq.
(\ref{27singlet}) interact with the adjoint sources again and a singlet is produced
according to Eq. (\ref{14singlet}).

It is observed from Fig.(\ref{all}) that the potentials are concave at
intermediate distances. It is worse for representation $14$. To get a
nonconcave potential and to be able to calculate the string tensions more
accurately, we use the method of  Ref. \cite{Rafi} which considers some random
fluctuations for the vortex profiles. Figure (\ref{short-dist}) shows the potentials
at intermediate distances.
A linear regime is observed for all three representations. However, the length of
this area is very small.
The Casimir scaling regime is expected to extend roughly from the onset of
confinement to the onset of screening. String tensions are calculated for $R$
between $5$ and $12$. From Fig. (\ref{short-dist}), the string tension ratios are
\begin{equation}
\frac{K_{14}}{K_{f}}=1.48, ~~~~~~~~~~~~~~~~~~~~~~~\frac{K_{27}}{K_{f}}=1.65,
\label{string ratio}
\end{equation}
while the Casimir ratios are
\begin{equation}
\frac{C_{14}}{C_{f}}=2, ~~~~~~~~~~~~~~~~~~~~~~~\frac{C_{27}}{C_{f}}=\frac{7}{3}.
\end{equation}
The string tension ratios are qualitatively in rough agreement with Casimir ratios.
The agreement between Casimir ratios and string tension ratios obtained from the
lattice calculations is very good \cite{Olejnik2008}. However, we recall that the
thick center vortex model predictions for proportionality with  Casimir scaling for
SU($2$), SU($3$), and SU($4$) have also been very rough \cite{Fabe1998,Deld}. The
Casimir ratios of representations $27$ to $14$ is $1.16$, which is in good
agreement with the string tension ratios obtained from the model, about
$1.12$, in Eq. (\ref{string ratio}). It is also possible to find the
potentials between static sources in other higher representations of the G($2$)
gauge group, but finding the Cartan subalgebra from the weight diagrams  would be a
little bit harder. However, studying these three representations of the G($2$) gauge
group shows that the thick center vortex model, with the idea of the domain structure,
works rather w
 ell even in the G($2$) gauge group which has only a trivial center.

\section{conclusion}

Studying confinement in gauge groups without nontrivial center elements is an
attractive subject. G($2$) is one of these exceptional groups which has only a
trivial center. Lattice calculations show some evidence for the confinement of
quarks at intermediate distances. On the other hand, the thick center vortex model
is a phenomenological model which gives potentials between quarks in gauge groups
with nontrivial center elements. In this paper, to reproduce the lattice results,
the idea of the vacuum domain structure has been used in the thick center vortex
model. A flux has been assigned to the trivial center of the G($2$), gauge group and
the potential between two static G($2$) quarks is obtained.  The potential is
screened at large distances, as expected from the thick center vortex model. In other
words, the trivial center does not affect the Wilson loop, where a complete link is
established between them  at large distances. In this regime, the vortex (domain)
 is located completely inside the minimal area of the loop. A linear potential at
intermediate distances is observed. The thickness of the domains and possibly the
SU($3$) subgroups of the G($2$) gauge group are responsible for this linear
behavior at intermediate distances.

In addition, potentials between static sources of the adjoint and the $27$
-dimensional representations are calculated by the model. In both cases, the
potentials are linear at intermediate distances, and the string tensions are
qualitatively in rough agreement with Casimir scaling.

\section{\boldmath Acknowledgments}

We would like to thank Manfried Faber and Stefan Olejnik for very helpful
discussions. We are grateful to the research council of the University of Tehran for
supporting this study.

\end{document}